# The same frequency of planets inside and outside open clusters of stars

Søren Meibom[1], Guillermo Torres[1], Francois Fressin[1], David W. Latham[1], Jason F. Rowe[2], David R. Ciardi[3], Steven T. Bryson[2], Leslie A. Rogers[4], Christopher E. Henze[2], Kenneth Janes[5], Sydney A. Barnes[6], Geoffrey W. Marcy[7], Howard Isaacson[7], Debra A. Fischer[8], Steve B. Howell[2], Elliott P. Horch[9], Jon M. Jenkins[10], Simon C. Schuler[11], Justin Crepp[12]

[1] Harvard-Smithsonian Center for Astrophysics, Cambridge, MA 02138 USA.

[2] NASA Ames Research Center, Moffett Field, CA 94035, USA.

[3] NASA Exoplanet Science Institute / California Institute of Technology, Pasadena, CA 91125, USA

[4] California Institute of Technology, Pasadena, CA 91125, USA

[5] Boston University, Boston, MA 02215, USA

[6] Leibniz-Institute for Astrophysics, Potsdam, Germany / Space Science Institute, USA

[7] University of California Berkeley, Berkeley, CA 94720, USA

[8] Yale University, New Haven, CT 06520, USA

[9] Southern Connecticut State University, New Haven, CT 06515, USA

[10] SETI Institute / NASA Ames Research Center, Moffett Field, CA 94035, USA

[11] National Optical Astronomy Observatory, Tucson, AZ 85719, USA

[12] University of Notre Dame, Notre Dame, IN 46556, USA

**Most stars and their planets form in open clusters. Over 95 per cent of such clusters have stellar densities too low (less than a hundred stars per cubic parsec) to withstand internal and external dynamical stresses and fall apart within a few hundred million years[1]. Older open clusters have survived by virtue of being richer and denser in stars (1,000 to 10,000 per cubic parsec) when they formed. Such clusters represent a stellar environment very different from the birthplace of the Sun and other planet-hosting field stars. So far more than 800 planets have been found around Sun-like stars in the field[2]. The field planets are usually the size of Neptune or smaller[3–5]. In contrast, only four planets have been found orbiting stars in open clusters[6–8], all with masses similar to or greater than that of Jupiter. Here we report observations of the transits of two Sun-like stars by planets smaller than Neptune in the billion-year-old open cluster NGC6811. This demonstrates that small planets can form and survive in a dense cluster environment, and implies that the frequency and properties of planets in open clusters are consistent with those of planets around field stars in the Galaxy.**

Previous planet surveys in clusters have suffered from insufficient sensitivity to detect small planets, and from sample sizes barely large enough to find the less common larger planets[9]. The recent discovery by the Doppler method of two giant planets around Sun-like stars in the Praesepe open cluster[8] set a preliminary lower limit to the rate of occurrence of hot Jupiters in that cluster. This frequency is not inconsistent with that in the field, after accounting for the enriched metallicity of Praesepe[10] and the positive correlation between stellar metallicity and the frequency of giant planets[11]. However, it does not address the frequency of smaller planets such as those more commonly found around field stars. NASA's Kepler telescope is sensitive enough to detect planets of the size of Neptune or smaller, using the transit technique.

Our detection of two mini-Neptunes (two to four Earth radii, $R_\oplus$) in NGC6811 is the result of a survey of 377 stars in the cluster as part of The Kepler Cluster Study[12]. The two planets, Kepler-66b and Kepler-67b, have radii of $2.8R_\oplus$ and $2.9R_\oplus$ and are each transiting (passing in front of) a Sun-like star in NGC6811 once every 17.8 and 15.7 days, respectively. Kepler-66b and Kepler-67b are the smallest planets to be found in a star cluster, and the first cluster planets seen to transit their host stars, which enables the measurement of their sizes.

The properties derived for the two planets depend directly on the properties determined for their parent stars (Kepler-66 and Kepler-67). Because the members of NGC6811 form a coeval, co-spatial and chemically homogeneous collection of stars, they trace a distinct sequence in the colour–magnitude diagram (Fig. 1a). This allows both their commonly held properties (such as age and distance) and their individual physical characteristics (such as masses, radii and temperatures) to be determined reliably from stellar evolution models[13,14]. Kepler-66b and Kepler-67b therefore join a small group of planets with precisely determined ages, distances and sizes. Table 1 lists the model-derived properties of the two planets and their host stars. Figure 1a shows the locations of Kepler-66 and Kepler-67 in the colour–magnitude diagram for NGC6811, and Fig. 2 displays their phase-folded transit light curves reduced and calibrated by the Kepler pipeline[15].

The membership of Kepler-66 and Kepler-67 to NGC6811 was established from a five-year radial-velocity survey (see Supplementary Information below). They are both secure radial-velocity members of NGC6811 and are located squarely on the cluster sequence in the colour–magnitude diagram (Fig. 1a). Their rotation periods listed in Table 1 were determined from the periodic, out-of-transit, brightness variations in the Kepler light curves, caused by star spots being carried around as the star spins (see Supplementary Information). The rotation periods provide additional confirmation of cluster membership, as they obey the distinct relationship between stellar rotation and colour observed for other members of NGC6811. Figure 1b shows the colour versus rotation period diagram plotted for radial-velocity members of the cluster[16].

Because of the large distance to NGC6811, the two host stars are too faint (see Table 1) for their radial velocities to be measured with sufficient precision to confirm the status of Kepler-66b and Kepler-67b as true planets in the usual way, that is, by establishing that their masses are in the planetary range. To validate them as planets we instead applied a statistical procedure known as BLENDER (see Supplementary Information), by which we have demonstrated that they are much more likely to be planets than false positives. We determined probabilities of only 0.0019 and 0.0024 that Kepler-66b and Kepler-67b are false positives.

To establish whether finding two mini-Neptunes in NGC6811 is consistent with the rate of occurrence of planets in the field, we conducted a Monte Carlo experiment using the known spectral type and magnitude distributions of the 377 member stars. We simulated true planets adopting distributions of planet sizes and orbital periods corresponding to those found in the Kepler field, along with planet occurrence rates based on a statistical study of the Kepler candidates that accounts for the incidence of false positives as well as incompleteness[5]. We retained only the simulated planets that would be detectable by Kepler on the basis of real noise estimates for each star. We repeated the simulation 1,000 times to predict the average number of transiting planets of all sizes we would expect to detect among the known cluster members observed by Kepler, as well as their period and size distributions (Fig. 3). The result, 4.0±2.0 planets, is consistent with our two planet detections. The expected number of 2.2±1.5 mini-Neptunes is also consistent with our detection of two such planets, and the lack of smaller and larger transiting planets in NGC6811 similarly agrees with their predicted detection rates of 1.2±1.1 for Earths and super-Earths (0.8–2$R_\oplus$) and 0.6±0.6 for giant planets (>4$R_\oplus$). Together, the results imply that the planet frequency in NGC6811 is consistent with that of the field.

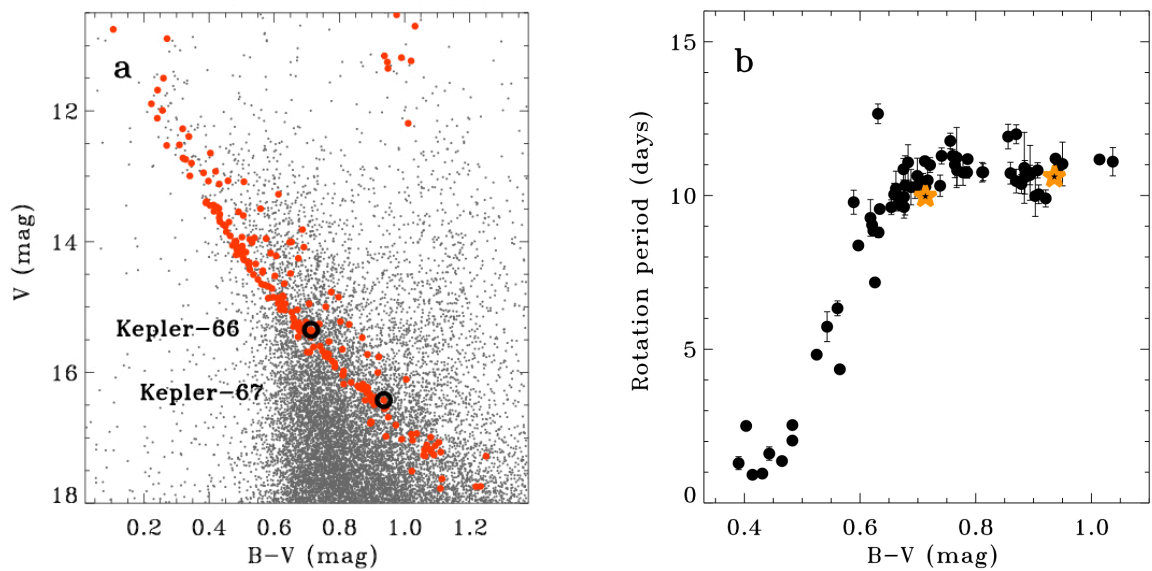

**Figure 1 | The color-magnitude and color-period diagrams for NGC6811. a)** The colour–magnitude diagram for stars within a 1-degree- diameter field centred on NGC6811 with the locations of Kepler-66 and Kepler- 67 marked by black circles. Cluster members, marked with larger red dots, trace a well-defined relationship between stellar mass (colour, B-V) and luminosity (brightness, V) that can be fitted by stellar evolution models to determine the age and distance of NGC6811 as well as the masses and radii of its members. By this method NGC6811 is found to be 1.00±0.17 billion years old and 1,107±90 parsecs distant[17]. **b)** The colour–period diagram for 72 NGC6811 members[16]. The rotation periods are determined from periodic brightness variations in the Kepler light curves, and the error bars represent the dispersion of multiple period measurements. As in the colour–magnitude diagram, cluster members trace a well-defined relation between stellar colour and rotation period. The locations of Kepler-66 and Kepler-67 on the cluster sequence are marked by orange star symbols.

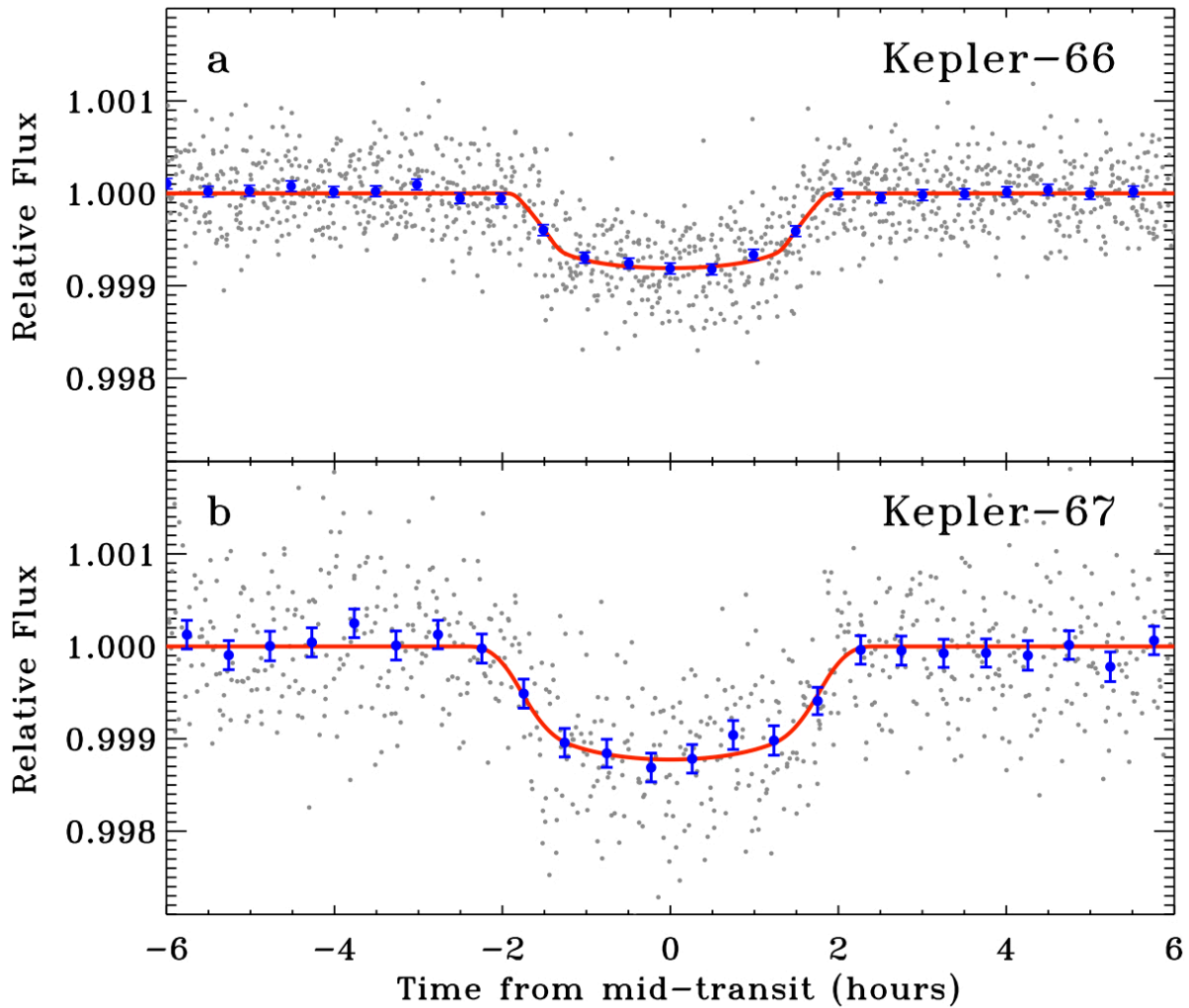

**Fig. 2 | Transit light curves.** The Kepler light curves for Kepler-66 (**a**) and Kepler-67 (**b**). The photometric measurements (grey points) were acquired in long cadence mode (30-min total exposures) and have been detrended[28], normalized to the out-of-transit flux level, and phase-folded on the periods of the transiting planets. The blue data points and error bars represent the same data phase-binned in 30-min intervals and the standard error of the mean, respectively. Transit models smoothed to the same cadence are overplotted in red.

The members of NGC6811 fall entirely within the range of stellar spectral types selected for the Kepler planet survey, and the slightly sub-solar metallicity of NGC6811 (ref. 17) is close to the average metallicity of the Galactic disk population from which the Kepler targets are drawn. Therefore, correlations between planet frequency and stellar mass and/or metallicity are not a concern when comparing the frequency and size distribution of planets in NGC6811 to that of the field. The detection of Kepler-66b and Kepler-67b thus places the first robust constraint on the frequency of small planets in open clusters relative to the field.

The comparison in Fig. 3 of the orbital periods and radii of Kepler- 66b and Kepler-67b with those in our simulated distributions shows that the sizes and orbital properties of the two planets are similar to those of the most common types of field planets (2–3$R_\oplus$, and orbital periods between 10 and 20 days). This suggests that the sizes and orbital properties of planets in open clusters are also not unlike those in the field.

The masses, structures and compositions of Kepler-66b and Kepler-67b can be constrained using theoretical models. With radii in excess of 2$R_\oplus$, the two planets probably contain significant quantities of volatiles in the form of astrophysical ices and up to a few per cent of H or He by mass. Volatile-poor rocky planets this large would have Saturn-like masses of 82–117 Earth masses (assuming an Earth-like composition with 32% iron core and 68% silicates by mass), and would be larger and more massive than any rocky exoplanet discovered to date. Instead, Kepler-66b and Kepler-67b are likely to have structures and compositions that resemble that of Neptune and, following mass–radius relations for exoplanets in the field[18], probably have masses less than 20 Earth masses (see Supplementary Information).

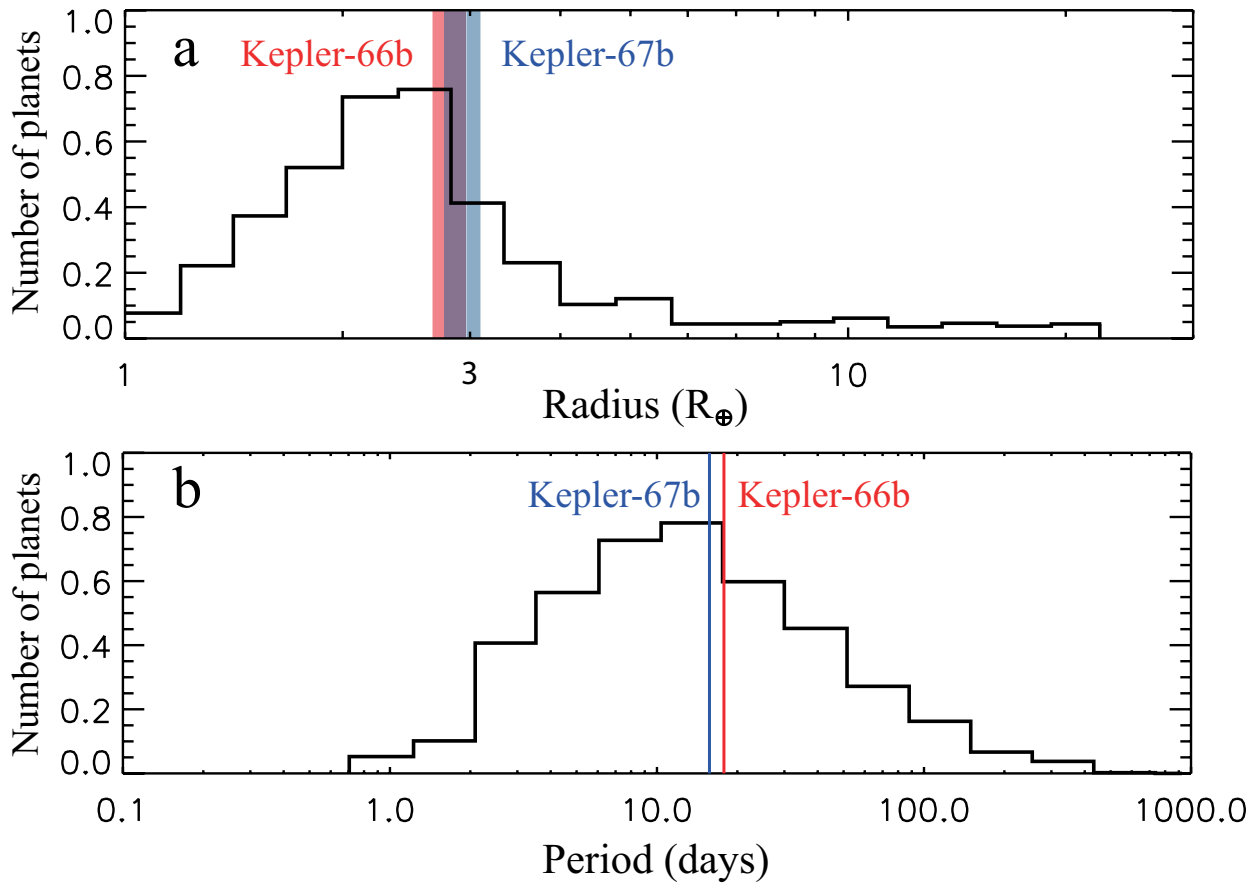

**Figure 3 | Distribution of planetary properties.** Histograms of planetary radii (**a**) and orbital periods (**b**) of simulated transiting planets expected in NGC6811, accounting for incompleteness and assuming the same period and size distribution and occurrence rate as in the field[5]. The properties of Kepler-66b and Kepler-67b are similar to those of the most commonly expected planets. The widths of the red and blue vertical lines reflect ±1σ errors in the radii and periods of the two planets.

For NGC6811 to have survived a billion years, the initial number density of stars in the cluster must have been at least that of the Orion Trapezium cluster (about 13,000 per cubic parsec) and thus more than two orders of magnitude greater than that of the typical cluster formed in a molecular cloud (about a hundred stars per cubic parsec; ref. 1). Highly energetic phenomena including explosions, outflows and winds often associated with massive stars would have been common in the young cluster. The degree to which the formation and evolution of planets is influenced by a such a dense and dynamically and radiatively hostile environment is not well understood, either observationally or theoretically[19–25]. The formation of planets takes place in the circum-stellar disks during the first few million years of a star's life, which is the typical lifetime of disks[26]. We estimated the number and mass-distribution of stars in NGC6811 at the time Kepler-66b and Kepler-67b formed by fitting a canonical initial mass function[27] to the current distribution of masses for members in the cluster (see Supplementary Information). The calculation suggests that the cluster contained at least 6,000 stars during the era of planet formation, including several O stars (masses greater than 20 solar masses) and more than one hundred B stars (masses between 3 and 20 solar masses). The discovery of two mini-Neptunes in NGC6811 thus provides evidence that the formation and long-term stability of small planets is robust against stellar densities that are extremely high for open clusters, and the violent deaths and high-energy radiation of nearby massive stars.

**Acknowledgements** Kepler was competitively selected as the tenth Discovery mission. Funding for this mission is provided by NASAs Science Mission Directorate. S.M. acknowledges support through NASA grant NNX09AH18A (The Kepler Cluster Study) and from the Kepler mission via NASA Cooperative Agreement NCC2-1390. G.T. acknowledges support through NASA's Kepler Participating Scientist Program grant NNX12AC75G.

**Author contributions**
S.M. is the PI of The Kepler Cluster Study and led the writing of the paper and the effort to identify members of NGC6811. He worked with G.T. and F.F. on characterization and validation of Kepler-66b and Kepler-67b, and with K.J. and S.A.B. on determination of the properties of NGC6811. G.T. developed the BLENDER software used to validate the planets, and determined the stellar properties of the host stars. F.F. worked on the BLENDER validation of the two planets and the Monte-Carlo simulation of the cluster yield. D.W.L. contributed follow-up spectroscopy of host stars. J.F.R. performed the light curve analysis to extract the planet characteristics. D.R.C. provided constraints on angular separation of potential background blends from adaptive optics imaging. S.T.B. performed pixel level centroid analysis. C.E.H. assisted in running BLENDER on the NASA Pleiades supercomputer. L.A.R. modeled the planets' interior structure to constrain the range of possible masses and compositions. K.J. led the supporting photometric study from which the bulk properties of NGC6811 are derived. S.A.B. Participated in the acquisition of ground-based spectroscopic and photometric data on NGC6811. G.W.M, H.I. obtained and analyzed high-resolution Keck HIRES spectra of host stars used for the BLENDER analysis. D.A.F. analysed HIRES spectra using SME. S.B.H., E.P.H obtained and analyzed speckle observations of the host stars. J.M.J. led the efforts of data collection, data processing, and data review that yielded the Kepler time series photometry. S.C.S. did spectroscopic analysis of stellar members of NGC6811 to aid in the determination of cluster parameters including metallicity. J.C. obtained AO imaging observations.

**Author Information** Correspondence and requests for materials should be addressed to S.M. (smeibom@cfa.harvard.edu).

**Table 1 | Stellar and planetary parameters for Kepler-66 and Kepler-67.**

| Stellar properties | Kepler-66 | Kepler-67 |
|---|---|---|
| Spectral Type | G0V | G9V |
| Effective temperature, $T_{eff}$ (K) | 5962 ± 79 | 5331 ± 63 |
| Log surface gravity (cm/s$^2$) | 4.484 ± 0.023 | 4.594 ± 0.022 |
| Rotation period (days) | 9.97 ± 0.16 | 10.61 ± 0.04 |
| Mass (solar masses) | 1.038 ± 0.044 | 0.865 ± 0.034 |
| Radius (solar radii) | 0.966 ± 0.042 | 0.778 ± 0.031 |
| Density (solar) | 1.15 ± 0.15 | 1.89 ± 0.17 |
| Visual (V) magnitude | 15.3 | 16.4 |
| Age (billion years) | 1.00±0.17 | |
| Distance (parsec) | 1107±90 | |
| Metallicity (Z) | 0.012±0.003 | |

| Planetary parameters | Kepler-66b | Kepler-67b |
|---|---|---|
| Orbital period (days) | 17.815815 ± 0.000075 | 15.72590 ± 0.00011 |
| Impact parameter | 0.56 ± 0.26 | 0.37 ± 0.21 |
| Time of mid transit (BJD) | 2454967.4854 ± 0.0025 | 2454966.9855 ± 0.0048 |
| Planet/star radius ratio | 0.02646 ± 0.00097 | 0.03451 ± 0.0013 |
| Scaled semi-major axis (a/$R_{star}$) | 30.3 ± 1.0 | 32.4 ± 1.1 |
| Semi-major axis (AU) | 0.1352 ± 0.0017 | 0.1171 ± 0.0015 |
| Radius ($R_⊕$) | 2.80 ± 0.16 | 2.94 ± 0.16 |

The age, distance and chemical composition of NGC6811 were determined from a maximum-likelihood fit of stellar evolution models[13,14] to the cluster sequence in the colour–magnitude diagram using Bayesian statistics and a Markov-chain Monte Carlo algorithm[17]. The best-fitting stellar isochrone[14] and photometric measurements in all available bandpasses (UBV, griz, JHK and D51 magnitude) were used to derive the effective temperatures, surface gravities, masses, radii and densities for Kepler-66 and Kepler-67. The transit and orbital parameters (period, impact parameter, time of mid-transit, radius ratio, and scaled semi-major axis) for Kepler-66b and Kepler-67b were derived from the Kepler photometry using a Markov-chain Monte Carlo procedure with the mean stellar density as a prior[28]. The parameters for Kepler-67b account for minor dilution from a close companion to the star described in section 3.2 of the Supplementary Information. Errors given for stellar and planetary parameters are 1σ uncertainties. BJD is barycentric Julian date, and AU is astronomical units.

# Supplemental Information

### 1. Kepler-66 and Kepler-67: cluster membership and rotation rates

The common space motion of the stars in a cluster is an effective way to distinguish them from foreground or background stars in the Galactic disk. Using the Hectochelle multi-object spectrograph on the MMT 6.5 m telescope, we have measured radial (line-of-sight) velocities over 5 years for more than 3,500 stars within a circular 1-degree diameter field centered on NGC 6811. With Hectochelle we observe in the spectral range from 5,150–5,300 Å with a spectral resolution of $\sim$ 40,200. For late-type stars like Kepler-66 and Kepler-67 this wavelength range is rich with narrow absorption lines and thus well suited for radial-velocity (RV) measurements. Our RV measurement precision for stars of F, G, and K type is $\sim 0.3\,\mathrm{km\,s^{-1}}$ down to 18.5 magnitude in $V$.

Against the broad velocity distribution of Galactic field stars, the members of NGC 6811 populate a distinct peak with a mean RV of $+7.7 \pm 0.8\,\mathrm{km\,s^{-1}}$. The uncertainty represents the $1\sigma$ velocity dispersion among cluster stars caused by internal dynamics, binary orbital motions, and observational errors. For a given star, the probability for cluster membership ($P_{RV}$) is calculated from simultaneous fits of separate Gaussian functions to the cluster ($F_C$) and field ($F_F$) RV distributions. The probability is defined as the ratio of the cluster-fitted value over the sum of the cluster- and field-fitted values at the star's RV[29]:

$$P_{RV} = \frac{F_C(RV)}{F_C(RV) + F_F(RV)}$$

The mean RVs for Kepler-66 and Kepler-67 of $+7.8\pm0.2\,\mathrm{km\,s^{-1}}$ and $+8.1\pm0.2\,\mathrm{km\,s^{-1}}$, based on 8 and 7 RV measurements, respectively, correspond to membership probabilities of 84% and 81%. Probabilities of less than 100% are expected as some background and foreground field stars will have RVs near the cluster mean. However, such field stars most likely lie below (background) or above (foreground) the cluster sequence in the color-magnitude diagram (CMD; see Fig. 1a in main text).

Because a unique and well-defined relationship between stellar rotation rate and color (proxy for stellar mass) has been established for members of NGC 6811[18] (see Figure 1b in the main text), the rotation periods of Kepler-66 and Kepler-67 provide additional confirmation of their membership to NGC 6811. The rotation periods of $9.97 \pm 0.16$ days and $10.61 \pm 0.04$ days for Kepler-66 and Kepler-67 place them on the tight rotational sequence traced by cluster members in the NGC 6811 color-period diagram (CPD; Figure 1b in main text). Background and foreground field stars are likely to be older and thus rotate more slowly than members of NGC 6811, placing them above the cluster sequence in the CPD. The rotation periods for Kepler-66 and Kepler-67 were determined from the *Kepler* light curves using a periodogram analysis to detect periodic variability. Supplementary Figures S1 and S2 show representative sections of *Kepler* light curves and the resulting periodograms for Kepler-66 and Kepler-67, respectively. The rotation periods for Kepler-66 and Kepler-67 are consistent with the spectroscopically measured projected rotational velocities ($v \sin i$) and radii for the two stars within the uncertainties of the latter.

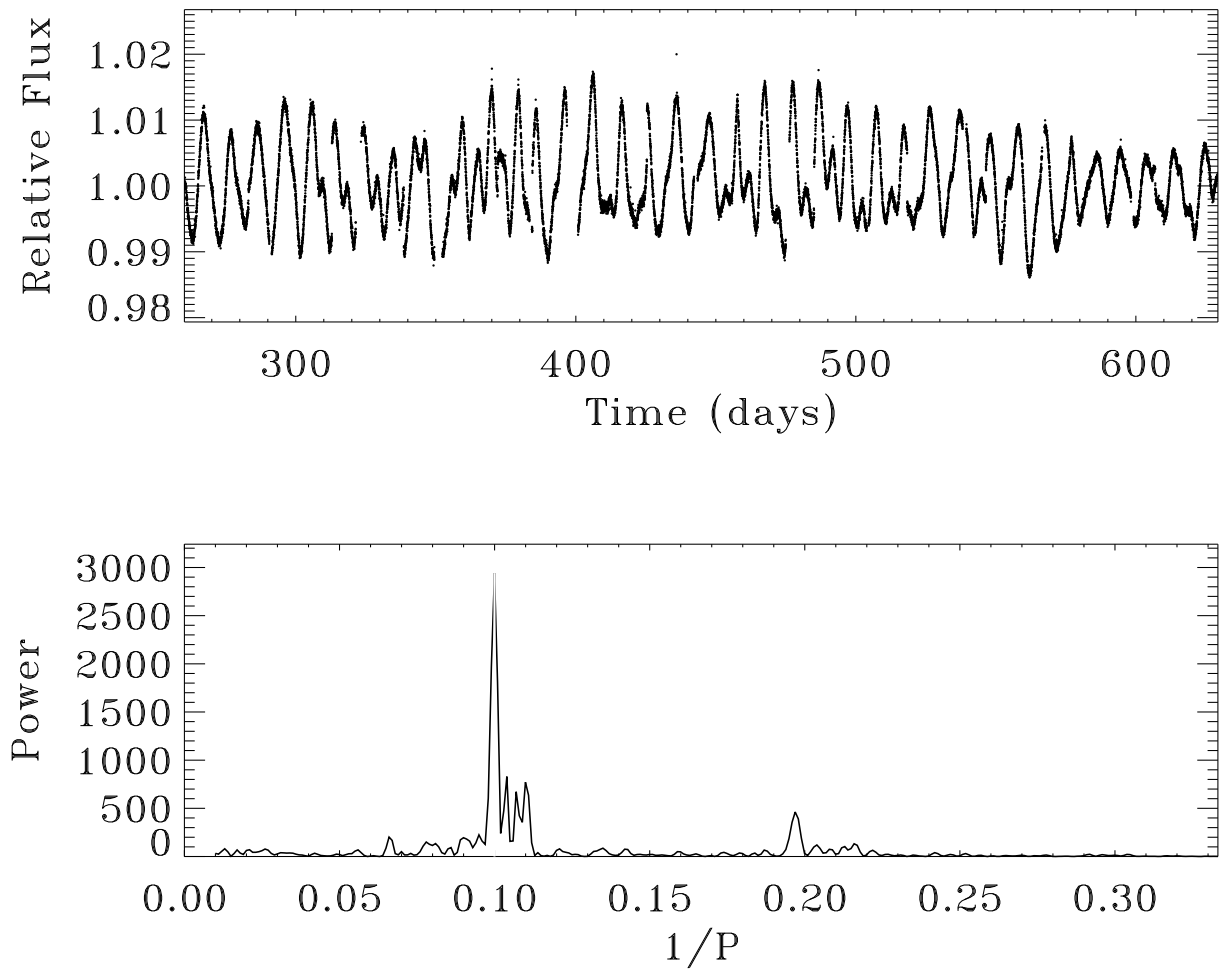

Fig. S1.— A segment of the *Kepler* light curve for Kepler-66 (top panel) and the corresponding periodogram (bottom panel). The power as a function of rotation frequency (inverse rotation period) in the periodogram is produced by the modulation by star spots of the stellar flux as the star rotates and peaks at a period of 9.97 days.

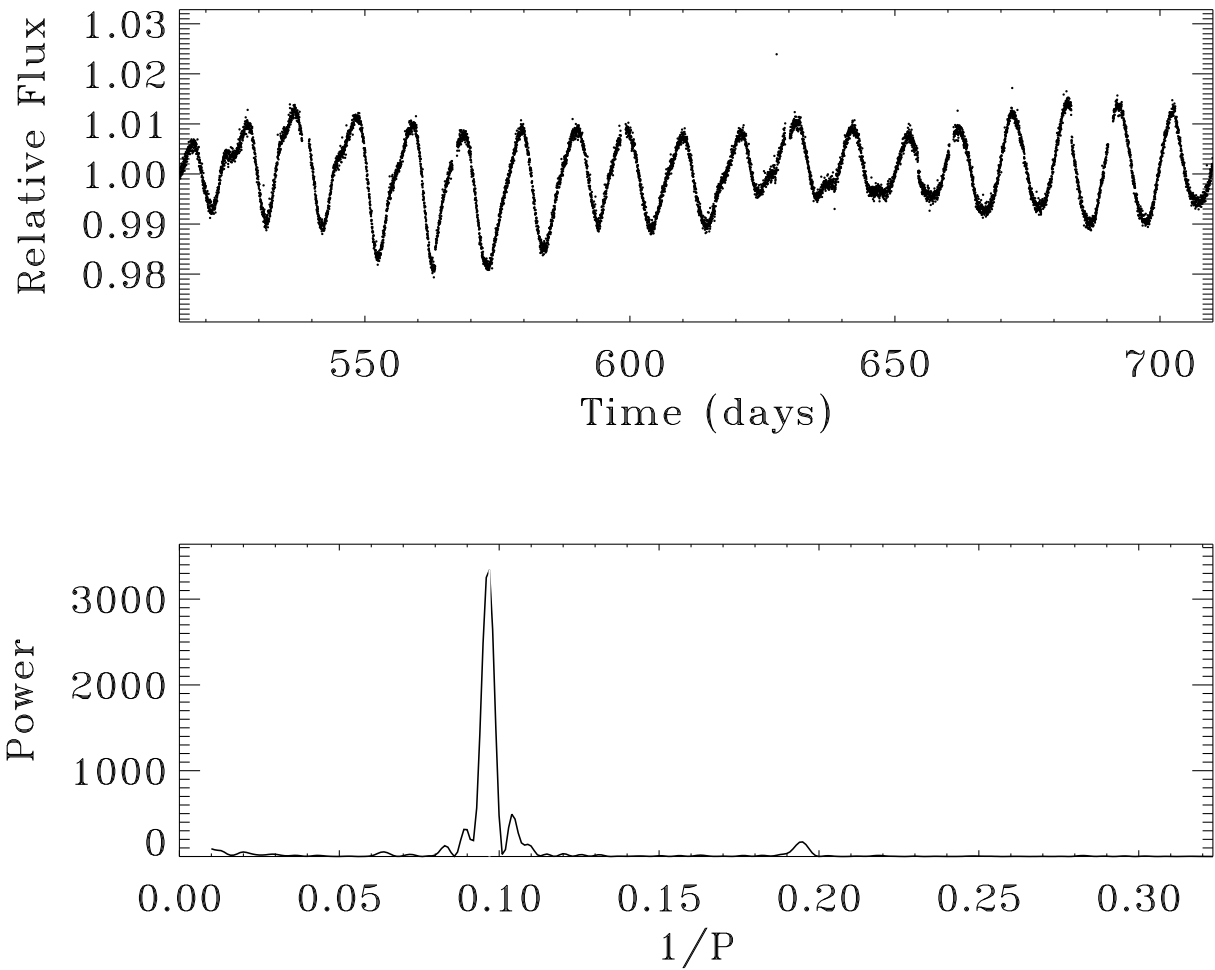

Fig. S2.— A segment of the *Kepler* light curve for Kepler-67 (top panel) and the corresponding periodogram (botom panel). The power as a function of rotation frequency (inverse rotation period) in the periodogram is produced by the modulation by star spots of the stellar flux as the star rotates and peaks at a period of 10.61 days.

## 2. Stellar population of NGC 6811 during the era of planet formation

Planets form in the disks of dust and gas surrounding stars for the first few million years of their lives. We can estimate the number of stars of different masses in NGC 6811 at the time Kepler-66 b and Kepler-67 b formed, by fitting a canonical initial mass function (IMF; [28]) to the current number of stars of mid-F spectral type ($M_\star \sim 1.3$–$1.4\,M_\odot$) in NGC 6811. We assume that stars in this mass-bin have been only minimally impacted by dynamical evolution in NGC 6811, as opposed to smaller stars that can be more easily ejected from the cluster after close encounters with more massive stars and binaries. Furthermore, the number of mid-F type stars in NGC 6811 should not be affected by stellar evolution as their main-sequence lifetimes are 4 to 5 times the cluster age. Nonetheless, the current number of mid-F type stars in NGC 6811 is likely lower than during the era of planet formation, making our estimates of the number of stars of different masses a lower limit. From the IMF fit we estimate that NGC 6811 contained at least 6000 stars, including eight O stars ($M_\star \gtrsim 20\,M_\odot$) and 125 B stars ($3\,M_\odot \lesssim M_\star \lesssim 20\,M_\odot$).

## 3. Validation of Kepler-66 b and Kepler-67 b as planets in NGC 6811

Transit signals can be mimicked by a variety of astrophysical phenomena unrelated to planets, such as a background or foreground eclipsing binary star or a star transited by a larger planet falling within the same photometric aperture as the target ('blends'). Analysis of the centroid motion in the *Kepler* images (Sect. 3.1) and analysis of additional high-spatial-resolution images (Sect. 3.2) show no such nearby stars that could cause the transit signals observed in Kepler-66 and Kepler-67. Low-precision radial velocities carried out with the MMT and Keck-I 10 m telescopes rule out stars or brown dwarfs as companions (Sect. 3.3). However, due to the faintness of the targets ($V = 15.3$ for Kepler-66 and $V = 16.4$ for Kepler-67) it is not practical to obtain the high-precision radial velocity measurements needed to detect the acceleration of the stars induced by a planet, which would be the customary way of confirming the planetary nature of these transit signals. We have therefore followed a statistical approach known as BLENDER [30–33] to show that the likelihood of a true planet transiting the target star far exceeds the likelihood of a false positive (Sect. 3.4).

### 3.1. Centroid motion analysis

To constrain the distance within which a background false positive could be the source of the transit we use a fit of the *Kepler* Pixel Response Function (PRF) to difference images. The PRF fit technique constructs difference images by subtracting averaged in-transit pixel values from out-of-transit pixel values. We compute the position of the *Kepler* PRF that best matches the difference image, giving the position of the star producing the transit signal. The PRF is also fit to the out-of-transit pixel image, giving the location of the target star assuming there are no other stars of comparable brightness in the out-of-transit pixel image. For further details see Bryson et al. (2013; [34]).

For Kepler-66 we find that the PRF fit of the transit source is offset from the PRF fit of the target star by $0\rlap.{''}14 \pm 0\rlap.{''}15$, or $0.93\sigma$. We rule out any star outside a $3\sigma$ radius of $0.44''$ as a potential source of a false positive. Centroid analysis of Kepler-67 is complicated by the existence of a brighter star (KIC 9532049, *Kepler* magnitude $Kp = 14.7$) about three pixels ($12''$) from the target star. The PRF-fit centroid of the out-of-transit image is dominated by this brighter star, and therefore is not a valid measurement of the target star's position. Instead we use the offset of the PRF-fit of the difference image from the KIC catalog position of the target star. This offset is $0\rlap.{''}60 \pm 0\rlap.{''}25$, or $2.37\sigma$. We rule out any star outside a $3\sigma$ radius of $0\rlap.{''}75$ as a potential source of a false positive.

### 3.2. High-resolution imaging

Near-infrared adaptive optics imaging of Kepler-66 and Kepler-67 were obtained on the nights of 29 May and 24 June 2012 with the Keck-II telescope and the NIRC2 near-infrared camera behind the natural guide star adaptive optics system. NIRC2, a $1024 \times 1024$ HgCdTe infrared array, was utilized in $0\rlap.{''}0099$ per pixel mode yielding a field of view of approximately $10''$. Observations were performed in the $K'$ filter ($\lambda = 2.124\,\mu\text{m}$; $\delta\lambda = 0.351\,\mu\text{m}$). The frames were dark-subtracted and flat-fielded into a final image for each filter.

The optical brightnesses of Kepler-66 and Kepler-67 are relatively faint ($Kp = 15.2$ and $Kp = 16.2$, respectively) making the adaptive optics correction difficult. As a result, the central cores of the resulting point spread functions (PSF) in the images have widths of FWHM $= 0\rlap.{''}2$ (approximately 20 pixels) for Kepler-66 and and FWHM $= 0\rlap.{''}7$ (approximately 70 pixels) for Kepler-67. No sources other than the primary target were detected in the field of view. Our point source detection limits were estimated in a series of concentric annuli drawn around the star. The separation and widths of the annuli were set to the FWHM of the primary target PSF. The standard deviation of the background counts is calculated for each annulus, and the $10\sigma$ limits are determined within annular rings[35]. The sensitivity curves for the $K'$ observations are shown in Supplementary Figure S3.

A wider companion to Kepler-67, 2 magnitudes fainter and falling just at the edge of the NIRC2 field at a distance of $4\rlap.{''}9$, was detected in *J*-band images from the United Kingdom Infra-Red Telescope (UKIRT) 3.8 m telescope. We account for the dilution caused by this contamination when computing the radius for Kepler-67 b reported in Table 1 of the main text. The effect is to increase the radius by about 7.7% compared to what it would be if we did not account for this companion.

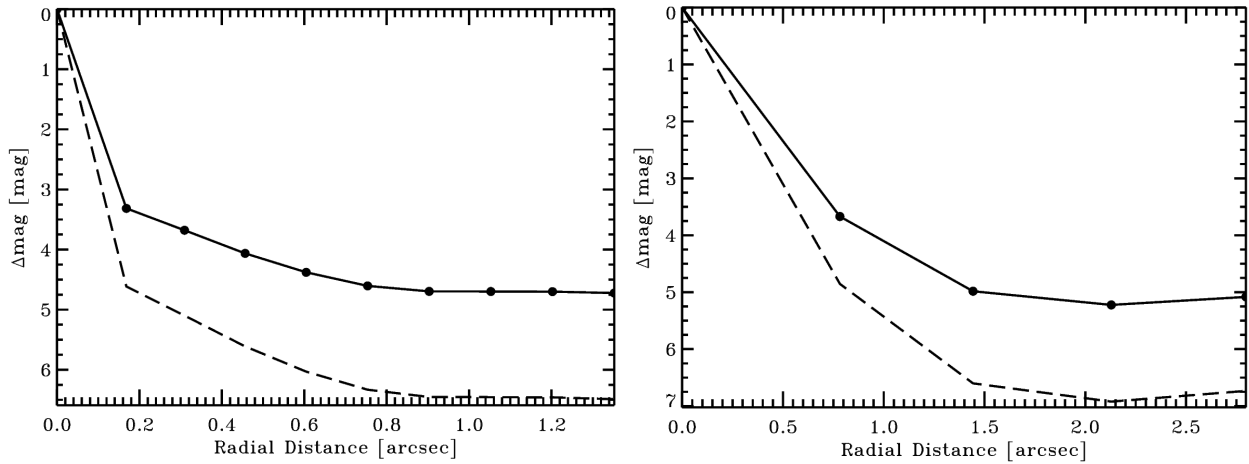

Fig. S3.— The point source sensitivity for the Keck-II NIRC2 $K'$ images of Kepler-66 (left) and Kepler-67 (right), as a function of angular distance from the target stars. The filled circles represent the sensitivity limits as measured in the $K'$ image in steps of the FWHM. The dashed lines represents the $K'$ limits converted to *Kepler* magnitudes based upon the expected colors of stars[35,36].

### 3.3. High-resolution spectroscopic observations

We used the HIRES instrument[37] on the Keck-I telescope (Mauna Kea, Hawaii) to obtain high-resolution spectra of Kepler-66 (2012 May 28 UT) and Kepler-67 (2012 June 22 UT) for the main purpose of placing limits on the presence of close stellar companions. The entrance slit used was $0''\!.87 \times 14''$, giving a resolving power of approximately $R = 55,000$. The signal-to-noise ratios per pixel for these spectra are 30 and 13, respectively, at a wavelength of 5,500 Å. To establish the sensitivity to companions we calculated the cross-correlation function (CCF) of the spectra based on all spectral orders between 4,900 Å and 6200 Å against a solar-type template, for which we used a spectrum of Ganymede obtained with the same instrumentation. Given that the two stars are not far from the Sun in spectral type, we expect the CCFs to be essentially symmetrical. Any lines of another star will typically be Doppler-shifted blueward or redward compared to the target, producing an asymmetry in the CCF. To better view such effects, we superimpose a left-right flipped version of the CCF, and the differences between the CCF and its mirror image enable us to infer upper limits on the brightness of companions. The CCFs of both stars have a single, strong, narrow peak with a full width at half maximum of some $20\,\mathrm{km\,s^{-1}}$. The differences with the mirror CCF are no larger than about 2%.

We conclude that there are no FGK stars within $0''\!.4$ of either target (half the width of the spectrometer slit) that are brighter than 2% of the primary star flux, corresponding to a magnitude difference of $\Delta Kp \sim 4$ mag. The only possible exception is a companion within $\sim 10 \, \text{km} \, \text{s}^{-1}$ of the RV of Kepler-66 or Kepler-67, which would be unresolved in the centroid motion analysis and the analysis of high spatial resolution imaging. If physically associated, any such companions would be orbiting beyond 10 AU of the star.

### 3.4. Validation with BLENDER

Here we seek to demonstrate that the odds ratio given by the expected frequency of planets of the sizes of Kepler-66 b and Kepler-67 b ('planet prior') divided by the frequency of blends is very large (several orders of magnitude). This same technique has been used previously to validate a number of *Kepler* candidates[36,38–40]. While those cases involved field stars, and chance alignments with other background or foreground stars in the field, for Kepler-66 and Kepler-67 some blends may result from chance alignments with stars that are within (i.e., members of) the cluster, *in addition* to those that may come from the field. This requires making assumptions about various properties of the cluster, whose stellar population represents an enhancement over the density of field stars in the direction of NGC 6811. We discuss this below.

Using BLENDER we simulated large numbers of background/foreground blend scenarios, as well as scenarios involving a smaller star or a planet eclipsing a physically associated star, and we compared the resulting light curves with the *Kepler* observations[1]. This allowed us to identify the range of blend parameters able to mimic the signals (i.e., the properties of the stars or planets involved, relative distances, etc.; see [33]). Many of these scenarios can be ruled out by limits on the presence of intruding stars available from our follow-up observations. These include the following: a) constraints from the centroid motion analysis, given in Sect. 3.1, excluding stars of any brightness beyond a certain angular separation; b) constraints from our high-resolution images in the form of sensitivity curves described and illustrated in Sect. 3.2, providing limits on the brightness of potential contaminating stars as a function of angular separation; c) limits on the brightness of closer companions from our high-resolution spectroscopic observations described in Sect. 3.3; and d) color information, which enables us to reject many false positives based on the fact that they would have an overall color inconsistent with the measured photometric indices available for the host star from our own photometric study of the cluster[13] and from the *Kepler* Input Catalog[41].

---

[1]In the case of Kepler-67, our simulations involved the extra dilution produced by the $\sim 5''$ companion reported in Section 3.2

For both Kepler-66 and Kepler-67 we found that all viable false positive scenarios involving eclipsing binaries blended with the target are at large distances behind the cluster. By viable blends we mean those that are not ruled out and that match the *Kepler* light curves within acceptable ($3\sigma$) limits[32]. Most scenarios involving a larger planet transiting another star blended with the target can be either behind or in front of NGC 6811. For all of these background/foreground blends (star + star, or star + planet) we determined the frequencies of false positives using informed estimates of star densities in the field[42] as well as rates of occurrence in the field of eclipsing binaries[43] and larger transiting planets involved in blends[5,44].

Other viable false positives involve planets transiting stars that are true members of NGC 6811, either physically associated with the target (in a hierarchical configuration) or not. For these we adopted binary properties from the work of Raghavan et al. (2010; [45])[2], and we require also estimates of the mean stellar density in the cluster (described below), and of the occurrence of planets *within* the cluster, which is of course not known *a priori*. As it turns out, however, the importance of blends coming from the cluster itself is relatively small (see below).

Estimating the planet prior for validation is the most delicate step of the procedure, as it also requires knowledge of the rate of occurrence of planets in the cluster, and is in fact directly proportional to that rate. In order to place rough limits on this unknown quantity that are sufficient for our purposes, we relied on the fact that we have found two candidates among the 377 member stars examined by *Kepler*. We then asked the following question: Given what we know about the sources of blends, how likely is it to find *two* false positives among these 377 stars (i.e., that *both* our candidates are blends)? To answer this question we performed a Monte Carlo simulation of blends for each of the 377 known members of NGC 6811 following the methodology of Fressin et al. (2013; [5]), adopting stellar properties for these stars as listed in the KIC. We counted how many false positives can mimic the light curves of small Neptunes (SN, with radii taken here to be between 2 and 4 $R_\oplus$) such as those implied by the Kepler-66 and Kepler-67 signals, and would also be detectable by *Kepler*. We repeated this simulation one million times to infer a probability. The result is only a 0.2% chance that we would have found two false positives. We then asked how likely it would be to find a single false positive, and a similar calculation yielded a 7.7% chance. We conclude that while we cannot rule out that one of our candidates is indeed a false positive, it is extremely unlikely that both are. In other words, at least one of our two candidates is a true planet. From this finding we may derive a lower limit to the frequency of detectable transiting small Neptunes in the cluster, equal to $f_{\rm SN}^{\rm cluster} = 1/377 = 2.7 \times 10^{-3}$.

---

[2]These binary properties refer strictly to field stars. The properties of binaries in NGC 6811 are not known, and we have assumed here that they are not appreciably different from the field. In particular, we assumed a frequency of multiple stars of 44%[45]. Adopting a multiplicity as high as 100% would not change our results below in any significant way.

It is of interest to compare this figure with the rate of occurrence of small Neptunes that *Kepler* is able to detect in the field. This may be estimated by simply counting the candidates of this size in the catalog of KOIs (*Kepler* Objects of Interest) by Batalha et al. (2013; [44]), and dividing by the total number of *Kepler* targets that are main-sequence stars ($N_{\rm targ} = 138{,}253$). Of the 985 KOIs of small-Neptune size, we expect a total of 66 to be false positives based on Monte Carlo simulations similar to those above. Their rate of occurrence in the field is then $f_{\rm SN}^{\rm field} = (985 - 66)/138{,}253 = 6.6 \times 10^{-3}$. Our rough estimate of the planet frequency in the cluster is therefore at least $2.7 \times 10^{-3}/6.6 \times 10^{-3} \approx 40\%$ of that in the field, for objects of this size.

Armed with this estimate, and the other ingredients mentioned earlier, we proceeded to compute the detailed blend frequencies for our two candidate transiting planets, to be compared later with the corresponding planet priors.

For Kepler-66, the expected frequency of blends involving background eclipsing binaries is $1.5 \times 10^{-8}$. Blends consisting of a larger planet transiting a background or foreground star have a frequency of $6.0 \times 10^{-7}$. The contribution from blends involving larger planets transiting cluster stars unrelated to the target was based on an estimate of the density of stars in NGC 6811 and the rate of occurrence of planets of suitable size. `BLENDER` indicates that in order to mimic the transits, such stars must all be of late spectral type (with masses between 0.25 and 0.45 $M_\odot$) and the planets involved in those blends are confined to a very narrow range of sizes. For computing the stellar density in the cluster within the small area of the sky allowed by the limits from our centroid motion analysis (Supplementary Sect. 3.1) we used a very conservative estimate of 1,000 M dwarfs in NGC 6811, and a cluster angular diameter of 1°. While in principle the rate of occurrence of planets to use for this particular source of false positives is the one in the cluster (for which we only have a lower limit, $f_{\rm SN}^{\rm cluster}$), it is conservative to adopt the larger rate for the field, $f_{\rm SN}^{\rm field}$, as this leads to a higher blend frequency. The result is a frequency of false positives involving cluster stars unrelated to the candidate of $4.3 \times 10^{-9}$, which is essentially two orders of magnitude smaller than the corresponding contribution from the field, demonstrating that blends of this kind within the cluster have a negligible impact. Finally, for the blends involving larger planets transiting a physical companion star to the target, we again used $f_{\rm SN}^{\rm field}$ instead of $f_{\rm SN}^{\rm cluster}$ to be conservative, and obtained a frequency of $8.5 \times 10^{-8}$. The sum of the four contributions yields a total expected blend frequency for Kepler-66 of $7.0 \times 10^{-7}$. The last two of the contributions just described, which correspond to blends involving larger planets transiting other stars within the cluster, represent only ∼13% of the total. Hence, changing the estimate of the frequency of planets in the cluster has relatively little influence on the result.

To arrive at the odds ratio for Kepler-66 we require an estimate of the expected frequency of true planets of this particular size in the cluster (planet prior), which we define to be those with a radius within a $3\sigma$ interval of the measured radius for this candidate. We obtained this prior by appealing once again to the KOI catalog of Batalha et al. (2013; [44]), keeping in mind that statistics derived from that list pertain strictly to the field, rather than the cluster, so a correction must be applied. We counted 393 KOIs that are similar in size to Kepler-66 b (within $3\sigma$ of its measured radius), of which we expect 28 to be false positives, from Monte Carlo simulations. Similar calculations indicate that such planets are only detectable by *Kepler* around 71% of its targets[5], which allows us to correct the census for incompleteness. The expected frequency of planets of this size in the cluster is then $(393-28)/(71\% \times 138{,}253) \times (f_{\rm SN}^{\rm cluster}/f_{\rm SN}^{\rm field})$, where the last factor adjusts the planet prior so that it corresponds to the cluster as opposed to the field.[3] Previously we had inferred a lower limit to this factor of 40%, implying an occurrence rate of small Neptunes in NGC 6811 that is no more than 2.5 times smaller than in the field. While we believe this to be a robust lower bound, we have chosen here to be more cautious, arbitrarily adopting a rate in the cluster ten times smaller than in the field for this calculation.[4] The result is a very conservative planet prior of $3.7 \times 10^{-4}$, which is 530 times larger than the total blend frequency given earlier. We therefore consider the Kepler-66 signal to be validated as due to a true planet to a very high degree of confidence (false alarm probability FAP = 0.0019).

Similar calculations for Kepler-67 yielded expected blend frequencies of $4.8 \times 10^{-7}$ (background eclipsing binaries), $6.5 \times 10^{-7} + 7.4 \times 10^{-9}$ (stars transited by a larger planet in the background/foreground of the cluster, and within the cluster but unrelated to the target), and $1.4 \times 10^{-7}$ (larger planets transiting a physical companion). The total frequency of false positives is $1.3 \times 10^{-6}$, of which only ∼11% come from larger planets transiting cluster members. The planet prior is $(571-41)/(71\% \times 138{,}253) \times (f_{\rm SN}^{\rm cluster}/f_{\rm SN}^{\rm field})$, and again using a planet rate of occurrence conservatively 10 times smaller than in the field, the result is $5.4 \times 10^{-4}$. Since this is 410 times larger than the total blend frequency (FAP = 0.0024), once again we obtain a clear validation of Kepler-67 b as a bona-fide planet in NGC 6811.

---

[3]The calculation of the planet prior detailed here is specific to Kepler-66, in the sense that the 393 KOIs, 28 false positives, and 71% completeness factor correspond strictly to a planetary radius interval within $3\sigma$ of that of Kepler-66 b. Therefore, an additional but reasonable assumption we make here is that the ratio $f_{\rm SN}^{\rm cluster}/f_{\rm SN}^{\rm field}$ for the broader category of small Neptunes ($2-4\,R_\oplus$) is not significantly different for planets in the more limited size interval just mentioned, which is likely to be approximately true given that the radius of Kepler-66 happens to lie in the middle of that range. Similarly for Kepler-67.

[4]Note that since the odds ratio we seek is defined as the planet prior divided by the total expected blend frequency, *decreasing* the estimated frequency of planets in the cluster relative to the field for the planet prior is conservative, as is *increasing* it for the blend frequency calculation described earlier.

## 4. Planet masses and structure

Despite the lack of dynamical confirmation of Kepler-66 b and Kepler-67 b, models of planet formation, structure, and survival can nonetheless yield useful constraints and insights about their masses and compositions.

Because the radii of Kepler-66 b and Kepler-67 b are in excess of 2 $R_\oplus$, it is likely that the planets contain significant quantities of volatiles in the form of H/He and/or astrophysical ices that contribute to the observed transit radii. Without volatiles, rocky planets large enough to match the Kepler-66b and Kepler-67b transit radii would need Saturn-like masses of 82 and 117 $M_\oplus$, respectively (based on Seager et al. (2007) mass-radius relations[46], assuming an Earth-like composition with 32% iron core and 68% silicates by mass). To date, all confirmed transiting planets with radii between 2 and 3 $R_\oplus$ have measured masses less than 20 $M_\oplus$ (see e.g., [19]).

We follow the modeling approach of Rogers et al. (2011; [47]) to constrain the masses and volatile contents of Kepler-66 b and Kepler-67 b. We first consider scenarios where Kepler-66 b and/or Kepler-67 b have H/He dominated atmospheres. Recent models of planet formation have indicated that low-mass (a few $M_\oplus$) proto-planets may plausibly accrete H/He gas from the protoplanetary nebula if formed in situ[48,49] and if formed beyond the snow-line[47]. If Kepler-66 b and Kepler-67 b have ice-less rocky interiors (formed inside the snow line), they can be no more than 4.3% and 6.2% H/He by mass, respectively (otherwise the planet transit radii would be larger than observed). If Kepler-66 b and Kepler-67 b instead have interiors comprised of a mixture of ice and rock (formed outside the snow line), H/He can account for no more than 1.2% and 2.5% of their masses.

An alternative scenario is one in which Kepler-66 b and Kepler-67 b formed beyond the snow line from a mixture of ice and rock, but did not manage either to accrete or to retain H/He to this day. In this case, a super-critical water-dominated envelope could account for the observed planet radii, provided the masses of Kepler-66 b and Kepler-67 b exceed 11 $M_\oplus$ and 15 $M_\oplus$, respectively.